

%

\input harvmac

\overfullrule=0pt


\def\r{\rho}
\def\a{\alpha}

\def\b{\beta}

\def\g{\gamma}

\def\d{\delta}
\def\D{\Delta}
\def\e{\epsilon}

\def\th{\theta}

\def\m{\mu}
\def\n{\nu}

\def\l{\lambda}
\def\s{\sigma}

\def\no{\noindent}

\def\bs{\bigskip}

\def\qq{\qquad}

\def\bl{\bigl}
\def\br{\bigr}


\def\IR{\relax{\rm I\kern-.18em R}}

\def\gh{$SL(2,\IR)_{-k'}\otimes SU(2)_k/(\IR \otimes \tilde \IR)$}




hep-th/9208xxx \hfill {USC-92/HEP-B3}
\rightline{August 1992}
\bs\bs

\centerline {$SL(2,\IR) \otimes SU(2)/ \IR^2$
            {\bf STRING MODEL IN CURVED SPACETIME } }
\centerline   {   {\bf AND EXACT CONFORMAL RESULTS }
                  {  \footnote{$^\dagger$}
 {Research supported in part by DOE, under Grant No. DE-FG03-84ER-40168. } }
               }

\vskip 1.00 true cm

\centerline {I. BARS and K. SFETSOS}

\bigskip

\centerline {Physics Department}
\centerline {University of Southern California}
\centerline {Los Angeles, CA 90089-0484, USA}


\vskip 1.50 true cm

\centerline{ABSTRACT}

Pursuing further the recent methods in the algebraic Hamiltonian approach
to gauged WZW models, we apply them to the bosonic \gh\ model recently
investigated by Nappi and Witten. We
find the global space and compute the conformally exact metric and dilaton
fields to all orders in the $1/k$ expansion. The semiclassical limit $k',k\to
\infty$ of our exact results agree with the lowest order perturbation
computation which was done in the Lagrangian formalism. We also discuss the
supersymmetric type-II and heterotic versions of this model and verify
the non-renormalization of $e^\Phi\sqrt{-G}$.

\vfill
\eject


The non-compact coset (or equivalently non-compact gauged WZW model) approach
to string theory in curved spacetime
\ref\IBhet{ I. Bars, Nucl. Phys. {\bf B334} (1990) 125. }
\ref\BN{ I. Bars and D. Nemeschansky, Nucl. Phys. {\bf B348} (1991) 89.}
has been under investigation during the past year, following the two
dimensional black hole interpretation of Witten
\ref\WIT{E. Witten, Phys. Rev. {\bf D44} (1991) 314.}.
Many of the later papers [4-13]
have used the gauged WZW Lagrangian approach
in a unitary gauge that yields an effective sigma model with a classical
background metric, antisymmetric tensor and one loop dilaton. In the patch of
the spacetime that is given by a particular unitary gauge, the interpretation
of the geometry ranged from black hole singularities to cosmology for various
cosets.
\nref\CRE{M. Crescimanno, Mod. Phys. Lett. {\bf A7} (1992) 489.}
\nref\HOHO{J. B. Horne and G. T. Horowitz, Nucl. Phys. {\bf B368} (1992) 444.}
\nref\BSthree{I. Bars and K. Sfetsos, Mod. Phys. Lett. {\bf A7} (1992) 1091.}
\nref\BSfour{I. Bars and K. Sfetsos, Phys. Lett. {\bf 277B} (1992) 269.}
\nref\FRA{E. S. Fradkin and V. Ya. Linetsky, Phys. Lett. {\bf 277B} (1992) 73.}
\nref\ISH{N. Ishibashi, M. Li, and A. R. Steif,
         Phys. Rev. Lett. {\bf 67} (1991) 3336.}
\nref\HOR{P. Horava, Phys. Lett. {\bf 278B} (1992) 101.}
\nref\RAI{E. Raiten, ``Perturbations of a Stringy Black Hole'',
         Fermilab-Pub 91-338-T.}
\nref\GER{D. Gershon, ``Exact Solutions of Four-Dimensional Black Holes in
         String Theory'', TAUP-1937-91.}
\nref\GIN{P. Ginsparg and F. Quevedo, ``Strings on Curved Space-Times:Black
         Holes, Torsion, and Duality'', LA-UR-92-640.}

In recent papers we showed how to improve on the perturbative Lagrangian
results by using algebraic Hamiltonian techniques to compute globally valid
\ref\BSglo{I. Bars and K. Sfetsos, ``Global Analysis of New Gravitational
Singularities in String and Particle Theories'', USC-92/HEP-B1
(hep-th/9205037), to appear in Phys. Rev. D (1992).}
and conformally exact
\ref\BSexa{I. Bars and K. Sfetsos, ``Conformally Exact Metric and Dilaton in
String Theory on Curved Spacetime'', USC-92/HEP-B2 (hep-th/9206006),
to appear in Phys. Rev. D (1992).}
\ref\KS{K. Sfetsos, ``Conformally Exact Results for
$SL(2,\IR)\otimes SO(1,1)^{d-2} /SO(1,1)$ Coset Models'',
USC-92/HEP-S1 (hep-th/9206048). }
geometrical quantities such as the metric and dilaton (and, in principle,
other fields) in gauged WZW models. We have applied the method to bosonic,
heterotic and type-II supersymmetric 4D string models that use the non-compact
cosets $SO(3,2)/SO(3,1)$, $\IR\times SO(2,2)/SO(2,1)$ and $\IR^2\times
SO(2,1)/\IR$
\foot{ A generalization of the coset $\IR\times SO(2,2)_{-k}/SO(2,1)_{-k}$
was given in \BN\ as
$\IR\times SL(2,\IR)_{-k_1}\times SL(2,\IR)_{-k_2}/SL(2,\IR)_{-k_1-
k_2}$ . The exact results for any $k_1,k_2$ may be deduced from
\BSexa\ with a little effort. Another model based on $\IR\times
SL(2,\IR)_{k_1}$ may be viewed as the limit $k_2\rightarrow\infty$ of the
generalized case, and therefore need not be considered separately.}.

There are just two remaining cosets that are relevant to strings on four
dimensional curved spacetime backgrounds. These are $SL(2,\IR)\times
SL(2,\IR)/\IR^2$ and $SL(2,\IR)\times SU(2)/\IR^2$ (for notational convenience
we do not distinguish between $\IR$ and $U(1)$ ). The latter coset was
investigated recently by Nappi and Witten
\ref\NAWIT{C. Nappi and E. Witten, ``A Closed, Expanding Universe in String
           Theory'', IASSNS-HEP-92/38.}
in the conformally perturbative semi-classical limit. In a particular patch
of the geometry they showed that the effective sigma model metric describes an
expanding and recollapsing universe. This coset was also examined before
\HOR\GER.
In the present note we reinvestigate this coset with the methods of
\BSglo\BSexa\ with the purpose of finding the global space, including all
dual patches, interpreting the geometry and obtaining the conformally exact
metric and dilaton fields. We will discuss the bosonic as well as
the heterotic and type-II supersymmetric versions of the string model. The
geometry for the $SL(2,\IR)\times SL(2,\IR)/\IR^2$ coset is obtainable just by
an analytic continuation from the one investigated presently, and therefore we
will not comment about it anymore in this paper.

It is convenient to parametrize the group element $g$ of $G$ as follows

\eqn\group{g=\pmatrix{g_1&0\cr0&g_2\cr}\ ,
\qq g_1 \in SL(2,\IR)\ ,\ \ g_2 \in SU(2)\ ,}

\no
with
\eqn\groupsl{g_1=\pmatrix{a&u\cr -v&b\cr}\ , \qq ab+uv=1}

\no
and
\eqn\groupsu{g_2=e^{i\g \s_2} e^{i s \s_3} e^{i \b \s_2}\ ,}

\no
where $\s_i,\ i=1,2,3$ are the standard $2\times 2$ Pauli matrices.
%
%
We will be interested below in the generators of left
(right) transformations $\d g=\e_L g$ ($\d g=g \e_R$). For $SL(2,\IR)$ they
are defined by $J_ig=-t^{(1)}_ig$ ($\bar J_ig=gt^{(1)}_i$) and for $SU(2)$
one has $I_ig=-t^{(2)}_ig$ ($\bar I_ig=gt^{(2)}_i$). The Pauli matrix
representations of the $SL(2,\IR)$ and $SU(2)$ generators are

\eqn\paulim{ t^{(1)}_1=\pmatrix{i{\sigma_1\over 2} & 0\cr 0& 0}, \qquad
t^{(1)}_2=\pmatrix{{\sigma_2\over 2} & 0\cr 0& 0}, \qquad
t^{(1)}_3=\pmatrix{i{\sigma_3\over 2} & 0\cr 0& 0}, \qquad
t^{(2)}_i=\pmatrix{ 0& 0\cr 0& {\sigma_i\over 2}}.        }

We gauge an abelian subgroup $H$ isomorphic to $\IR \otimes \IR$ generated
by the following infinitesimal transformations of $g_1$, $g_2$

\eqn\infin{\eqalign{&\d g_1=
\e {\s_3\over 2} g_1 +(\tilde \e \cos \a +\e \sin \a)g_1 {\s_3\over 2}\cr
&\d g_2=i\sqrt{{k'\over k}}\bl(\tilde \e {\s_2\over 2} g_2+
(-\tilde \e \sin \a +\e \cos \a)g_2 {\s_2\over 2} \br)\ .\cr}}
The relative coefficients in various terms have been fixed to insure
gauge invariance
\foot{There is a more general set of coefficients that are consistent with
gauge invariance, as follows: In the
transformation of $g_1$ replace $\cos\alpha$ with $\epsilon_1\cos\alpha$ and
in the transformation of $g_2$ replace $\cos\alpha$ by $\epsilon_2\cos\alpha$
and $\sin\alpha$ by $\epsilon_1\epsilon_2\sin\alpha$. The $\epsilon_i$ are
independently chosen as $\pm$. Such sign switches generate non-trivial
discrete duality transformations of the type discussed in \BSthree . These are
the analog of the vector/axial (or $R\rightarrow 1/R$) duality in the 2D black
hole model. In the discussion that follows we will mention the effect of these
signs on the geometry. }.
The generators of H may be identified by writing \infin\ in
the form $\d g=i(\e Q g +\tilde \e \tilde Q g)$,

\eqn\qqs{\eqalign{
&Q= J_3 -\sin \a\ \bar J_3 +\sqrt{k'\over k}\cos \a\ \bar I_2 \cr
&\tilde Q=- \sqrt{k'\over k} I_2 -\cos \a\ \bar J_3
- \sqrt{k'\over k}\sin\alpha\ \bar I_2 \ .\cr } }
Then it is easy to check that the condition for gauge invariance, or anomaly
cancellation
 \ref\anom{E. Witten, Comm. Math. Phys. {\bf 144} (1992) 189.} \BSthree\
 in the gauged WZW model is satisfied. Namely,

\eqn\ano{Tr\bl(Q_L^a Q_L^b\pmatrix{-k'&0\cr 0& k}\br)
     =Tr\bl(Q_R^a Q_R^b\pmatrix{-k'&0\cr 0& k}\br)\ , }
where $Q_L^a=(Q_L,\tilde Q_L)$ and $Q_R^a=(Q_R,\tilde Q_R)$ are the
matrix representations of $Q,\tilde Q$ applied on the
left and right of $g$, as follows from \paulim\qqs . The reason for the $k',k$
insertions is understood by writing the action in terms of traces over $g$
rather than $g_1,g_2$ individually.

The central charge of the model is

\eqn\central{c={3k'\over k'-2}+{3k\over k+2}-2\ .}
This model may be taken in only four dimensions by requiring the appropriate
central charge, as suggested in
\ref\IBhet{I. Bars, ``Heterotic String Models in Curved Spacetime", USC-
92/HEP-B4},
 or one can tensor this model with another conformal field theory representing
some internal space and demand that the total central charge be $c_{tot}=26$
(or $c_{tot}=15$ when supersymmetric). If, as in \NAWIT, one requires for the
curved spacetime model the same central charge as that of flat Minkowski
space, namely $c=4$, then \central\ gives the condition $k'=k+4$. This
particular choice of the central charge has the feature that both $k'$ and $k$
can continously reach infinite values and therefore the 1--loop results
obtained in \NAWIT\ correspond to a conformal model with the correct central
charge. Any other choice of $c$ requires finite $k'$ or $k$ even if one of
them is infinite, and therefore higher loop corrections are of substantial
importance.

With the Hamiltonian method (see \BSexa\ for more details)
the metric and dilaton for the coset
$G/H$=\gh\ are determined by comparing the two sides of the following equation

\eqn\lzero{(L_0+\bar L_0)\ T=-{1\over e^{\Phi}\sqrt{-G}}
\partial_\m \bl (G^{\m\n}e^{\Phi}\sqrt{-G}\partial_{\n}T\br )\ ,
       \qq \m,\n=0,1,2,3\ ,}

\no
where $L_0$, $\bar L_0$, the zero modes of the stress tensors for the
left and right movers respectively, are represented as second order
differential operators on group space, and $T$ is a tachyon level state which
is taken to be a gauge singlet. The appropriate algebraic expression for
$L_0$, which contains the exact dependence on $k$ and $k'$, is

\eqn\lzerol{L_0={\D_{SL(2,\IR)} \over -k'+2} +{\D_{SU(2)} \over k+2}-
{Q_L^2 \over k'} -{\tilde Q_L^2 \over k'}\ ,}

\no
where, as seen from \qqs , $(Q_L,\tilde Q_L)=(J_3,-\sqrt{k'/k}I_2)$ are the
zero modes of the left-moving subgroup currents whose central extensions are
$(k',k')$. The Casimir operators defined in terms of {\it hermitian} generators
are

\eqn\casil{ \D_{SL(2,\IR)}=-J_3 ^2 +{1\over 2}(J_+ J_- +J_- J_+)\ ,
\qq \D_{SU(2)}=I_1 ^2 +I_2 ^2 +I_3 ^2\ , }

\no
where $J_\pm=J_2\mp J_1$. The corresponding expressions for the right movers
are

\eqn\lzeror{\bar L_0={\bar \D_{SL(2,\IR)} \over -k'+2}
+{\bar \D_{SU(2)} \over k+2}-{ Q_R^2 \over k'} -{\tilde Q_R^2 \over k' }\ , }

\no
where, as seen from \qqs , the right moving current
$Q_R=(-\bar J_3 \sin\alpha
+\bar I_2 \sqrt{k'\over k}\cos\alpha)$ has central extension $k'\sin^2\alpha
+ k (\sqrt{k'/k}\cos\alpha)^2=k'$, and similarly $\tilde Q_R=(-\bar J_3
\cos\alpha -\bar I_2 \sqrt{k'/k}\sin\alpha)$ has central extension
$k'\cos^2\alpha + k(\sqrt{k'/k}\sin\alpha)^2=k'$. The formula for $\bar
L_0$ can be simplified by noticing that

\eqn\notice{-{Q_R^2 \over k'} -{\tilde Q_R^2 \over k' }=-{\bar J_3^2\over
k'}-{\bar I_2^2\over k}.   }
Thus, despite the complicated intermediate steps, $L_0$ and $\bar L_0$ end up
having the same structure in terms of the respective left and right moving
currents. Note also that the group Casimir operators are always the same for
the left and and right movers, i.e. $\bar \D_{SL(2,\IR)}=\D_{SL(2,\IR)}$ and
$\bar \D_{SU(2)}=\D_{SU(2)}$.

Now consider the differential operator form of the left and right generators
whose action on the group element reproduces the matrix representation of
\paulim . We take the {\it hermitian} $SL(2,\IR)$ generators in the form
given in \KS\ with an extra factor of $i$

\eqn\gensl{\eqalign{&J_3={i\over 2}(v \partial_v
-a \partial_a -u \partial_u)\ , \qq\cr
&J_+=i(b \partial_u -v \partial_a)\ ,\cr
&J_-=ia \partial_v\ ,\cr}
\eqalign{&\bar J_3={i\over 2}(a \partial_a
-u \partial_u +v \partial_v)\cr
&\bar J_+=-ia\partial_u \cr
&\bar J_-=i(u \partial_a -b\partial_v)\ .\cr}}

\no
In the above expressions the parameters $a$, $u$, and $v$ were taken as the
independent ones, whereas $b=(1-uv)/a$.
For the $SU(2)$ currents we have the following expressions

\eqn\gensul{\eqalign{
&I_1={i\over 2}\bl({\cos 2\g \over \sin 2s}\partial_{\b}
-\sin 2\g\ \partial_s -\cos 2\g \cot 2s\ \partial_{\g}\br)\cr
&I_2={i\over 2}\partial_{\g}\cr
&I_3={i\over 2}\bl({\sin 2\g \over \sin 2s} \partial_{\b}
+\cos 2\g\ \partial_s
-\sin 2\g \cot 2s\ \partial_{\g} \br)\ ,\cr}}

\no
and

\eqn\gensur{\eqalign{&\bar I_1={i\over 2}\bl(-\cos 2\b \cot 2s\ \partial_{\b}
-\sin 2\b\ \partial_s +{\cos 2\b \over \sin 2s}\partial_{\g}\br)\cr
&\bar I_2=-{i\over 2}\partial_{\b}\cr
&\bar I_3={i\over 2}\bl(\sin 2\b \cot 2s\ \partial_{\b}- \cos 2\b\ \partial_s
-{\sin 2\b \over \sin 2s}\partial_{\g}\br)\ .\cr}}

\no
The $SL(2,\IR)$ and $SU(2)$ Lie algebras are indeed obeyed by the right and
left generators separately and moreover any left generator commutes with any
right generator.

We are now prepared to consider the differential equation obeyed by a gauge
singlet state $T$ at the tachyon level. A priori $T(g_1,g_2)$ is a function
of the 6 group parameters $u,v,a,\beta,\gamma, s$. However, since it is
assumed to be a gauge singlet, it can be a function of only gauge singlet
combinations of these parameters, of which there are four, i.e. the four
dimensional spacetime. Then it automatically obeys the gauge invariance
conditions

\eqn\constr{ QT=0\ , \qq \tilde QT=0\ . }
{}From this one can also prove immediately that the condition
$(L_0 -\bar L_0)T=0$ for closed bosonic strings is satisfied.

In order to find the global geometry we must construct the 4 gauge invariant
combinations of group parameters that are identified with the 4D spacetime.
These are

\eqn\inva{s\ ,\ r=uv\ ,\ \r=\b+\g
-{1\over 2}\sqrt{{k'\over k}}{\cos \a \over 1+\sin \a}\ln {a\over b}\ ,\
\l=\g-\b+{1\over 2}\sqrt{{k'\over k}}
{\cos \a \over 1-\sin \a}\ln {u\over v}\ .}
where $b$ is to be expressed in terms of $a,u,v$ as above.
One can check that, these are indeed invariants by extracting the
infinitesimal transformation of each parameter from eq.\infin .


 When the Virasoro operators $L_0,\bar L_0$ are now applied on a function of
the 4 global string variables $X^\mu=(r,s,\rho,\lambda)$, i.e. $T(X)$, they
may be rewritten as second order differential operators in 4 dimensional
spacetime rather than the 6 dimensional group space,  by using the chain rule.
Then we are in a position to compare to the right hand side of \lzero\ and
extract the inverse metric $G^{\mu\nu}(X)$ from the double derivative terms
and the dilaton $\Phi(X)$ from the single derivative terms. We emphasize that
the exact dependence on $k,k'$ is included in these expressions. Finally, by
taking the inverse of $G^{\mu\nu}$ we derive the line element

\eqn\met{dS^2=2(k'-2) \bl (G_{rr} dr^2 +G_{ss} ds^2 +G_{\r\r} d\r ^2 +G_{\l\l}
d\l ^2\ \br )\ ,}

\no
with

\eqn\metri{\eqalign{&G_{rr}=-{1\over 4r(1-r)} \cr
&G_{ss}={k+2 \over k'-2} \cr
&G_{\r\r}={1+\sin\a \over 1- \sin \a} {1-r \over (1-r) B_s
+{k' \over k} r} \cr
&G_{\l\l}={1-\sin\a \over 1+ \sin \a} {r \over r A_s
+{k' \over k} (1-r) }\ , \cr } }

\no
where $A_s$, and $B_s$ are functions defined as follows

\eqn\asbs{\eqalign{&A_s={2\over k}+{1-\sin \a \over 1+\sin \a}{k'-2\over k+2}
(\cot ^2 s -{2\over k})\cr
&B_s={2\over k}+{1+\sin \a \over 1-\sin \a}{k'-2\over k+2}
(\tan ^2 s -{2\over k})\ .\cr } }

\no
The expression for the dilaton is

\eqn\dila{\Phi={1\over 2} \ln \bl[ \sin^2 2s
\bl(r A_s +{k'\over k} (1-r)\br) \bl((1-r) B_s +{k'\over k} r)\br) \br]
+  \Phi_0\ .}

Let us now take the large $k,k'$ limit and compare our global semi-classical
geometry to the one discussed in \NAWIT . In the global space the
semi-classical line element and dilaton take the form
\foot{Note that the combination $e^{\Phi} \sqrt{-G}$ is $k,k'$ independent.
Either the exact or the semi-classical metric and dilaton give the same
result. The non-renormalization of this quantity has been noticed for all
coset models \BSexa\KS .}

\eqn\classic{\eqalign {
 &dS^2=2k'{dr^2\over 4r(r-1)}+2k\bl (ds^2+{d\rho^2\over \tan^2s-{r\over r-1}
\tan^2({\pi\over 4}-{\alpha\over 2})} +{d\lambda^2\over \cot^2s-
{r-1\over r}\cot^2({\pi\over 4}-{\alpha\over 2})}\br) \ ,\cr
 & \Phi=\ln[r\ \tan^2({\pi\over 4}-{\alpha\over 2})\ \cos^2s -(r-1)\ \sin^2s] +
{\Phi}'_0\ .} }
The scalar curvature for the semi-classical geometry is

\eqn\curv{ \eqalign {R=
 & {7\over 4}{{4\over k'}r(1-r)(\cos 2s-\sin\alpha)^2-{1\over k}\sin^22s(1-
2r+\sin\alpha)^2 \over
    \bl (\sin^2s\ (1+\sin\alpha)+r\ (\cos 2s-\sin\alpha)\br )^2 } \cr
 & + 5 {{1\over k'}(1-2r)(\sin \a -\cos 2s)+{1\over k}\cos 2s\ (1-2r+\sin \a )
  \over \sin^2s\ (1+\sin\alpha)+r\ (\cos 2s-\sin\alpha) }+3({1\over k}-
 {1\over k'})\ . } }
{}From the definition \inva\ of the global variables, and the ranges of the
group parameters from which they are constructed, we deduce that $s$ is
periodic and taken in the range $0<s<\pi$, while $r,\rho,\lambda$ take values
from $-\infty$ to $+\infty$. In the range $0<r<1$ there are no singularities
in the metric, dilaton or scalar curvature. At generic values of $s$ there are
no curvature singularities at $r=0,1$ even though the metric appears
singular; this is simply a coordinate singularity which is eliminated by a
redefinition of variables. However, at $r=0$ and  $s=0,\pi$ the
dilaton and scalar curvature blow up. Similarly, at $r=1$ and $s=\pi/2$ these
quantities are singular. These points are shown by heavy dots on the figure.
 In addition, as may be seen from the metric, for
$r>1$ or $r<0$ there are curvature and dilaton singularities at arbitrary
values of $r$ when

\eqn\singular{s=\theta(\alpha,r),\quad {\rm or}\quad \pi-\theta(\alpha,r) }
where $\tan^2\theta(\alpha,r)={r\over r-1}\tan^2
 ({\pi\over 4}-{\alpha\over 2})$.
The singularities are shown as lines in the global $(r,s)$ space in Fig.1.
These lines approach asymptotically $s=s_0=\th(\alpha,\infty)$. On this figure
we have also indicated the signature in the basis $(r,s,\rho,\lambda)$. We see
that, as in other examples we studied there is a single time coordinate, but
the role of time switches from one coordinate to another in different patches.
Geodesics can cross the dotted boundary lines at $r=0,1,\infty$\ and $s=0,\pi$
(periodic), thus connecting all the regions into each other in principle.

The geodesics can be explicitly solved with the method described in \BSglo ,
but we will not discuss this point here in detail. However, let us mention
what happens when a geodesic touches one of the singularity lines. A
generic geodesic (time-like or light-like) seems to come tangentially to the
singularity and then bounce from it. But actually what it does is move into
another ``world" that is a copy of the original one and glued to it at the
singularity lines. It is described by the same invariant
coordinates $(r,s,\rho,\lambda)$, but has a different {\it discrete} gauge
invariant label. This corresponds to multicovers of the group space. This
phenomenon occurs in all previous models as well, even for the 2D black hole,
although it has not been discussed explicitly.  We do not have the space to
explain this phenomenon here, but simply say that it can be reconstructed from
the explicit geodesics for the entire group manifold (as in \BSglo\ ) and
cannot be seen by concentrating on only the 4 global invariants (or working
in specific gauges). The point is that there are additional discrete
gauge invariants that label the different ``worlds".

As in other non-compact coset models, there are interesting
discrete duality transformations \BSthree\BSfour\BSglo\ in the present model.
The dual models and patches are generated by the $\epsilon_1,\epsilon_2$ signs
mentioned in footnote 2. If these signs are inserted in the expressions for
$Q,\tilde Q$ in \qqs\ and in the invariants \inva\ then instead of \metri\
and \dila\ the dual metric and dilaton get generated. The result is the
following: in \metri\ and \dila\ replace $(s,\rho,\lambda)$ by $({\pi\over 2}-
s,\tilde \lambda,\tilde \rho)$, where $(\tilde\lambda,\tilde\rho)$ is a new
point in the $(\rho,\lambda)$ space
\foot{The $\epsilon_1,\epsilon_2$ dependence of these quantities are given by
$\r=\b+\e_1\e_2\g-{\e_2\over 2}\sqrt{{k'\over k}}{\cos \a \over 1+\sin \a}\ln
{a\over b}$ and $\l=\e_1\e_2\g-\b+{\e_2\over 2}\sqrt{{k'\over k}} {\cos \a
\over 1-\sin \a}\ln {u\over v}\ .$}.
Another way to view this result
is to leave the $(s,\rho,\lambda)$ unchanged, but instead interchange $r$ with
$(r-1)$ and also change the parameter $\alpha$ to $-\alpha$ . In the second
point of view the global $(r,s)$ space will be described by a figure obtained
from Fig.1 by a left-right reflection at $r={1\over 2}$ and by changing $s_0$
to ${\pi\over 2}-s_0$. So the dual regions can be obtained by comparing these
two figures.

In previous studies of the present coset \HOR\GER\NAWIT\ the choice of unitary
gauges forced the authors to concentrate on some of the patches in our figure.
In specific gauges it is not possible to recognize the full global space. To
see how this happens, let us consider one of the gauges in \NAWIT\ , namely
$a=b=\cos\psi$ and $u=v=\sin\psi$, so that the four independent parameters are
$r=uv=\sin^2\psi$, $\rho=\beta+\gamma$, $\gamma=\beta-\gamma$ and $s$.
Then our metric and dilaton in \classic\ become identical to those displayed
by Nappi and Witten \NAWIT\ when $k=k'$. We see that in this gauge we only
recover the patch labelled $II$ in Fig.1, and moreover $\rho,\lambda$ appear
to be periodic, although in the full global space this is not true. The second
gauge in \NAWIT\ parametrized by $x$ is also insufficient to reveal the
structure of Fig.1. Similar comments apply to the gauges studied in \HOR\GER .

We now move on to a brief discussion of the type-II supersymmetric and
heterotic versions of the string theory based on the present coset.
The algebraic structure of the supersymmetric coset and a description of the
action has been given elsewhere \IBhet . We wish to compute the conformally
exact metric and dilaton following the general discussion in \BSexa . The main
point is that the presence of the supersymmetric fermionic partners leads to a
modification of the gauge currents in such a way that they acquire new central
charges $k+2,\ k'-2$ that are shifted by the Coxeter numbers of the
corresponding groups. This has an important impact on the $k,k'$ dependence of
the Virasoro operators $L_0,\bar L_0$ as well as on the geometry that follows
from them through the Dalambertian.

We first explain how the shifting in the level occurs since there are some new
features not encountered before. The complications are due to the fact that
there are two different central extensions and the gauge currents are
constructed by mixing currents that have these different extensions.
For definiteness we consider the left movers which are assumed to be
supersymmetric. If the right movers are also supersymmetrized (as in the
type-II superstring) the same discussion will apply to them. We introduce
left-moving fermions classified in $G/H$. This means $SL(2,\IR)/\IR$ fermions
$\psi_1,\psi_2$ and $SU(2)/U(1)$ fermions $\chi_1,\chi_3$. The left-moving
part of gauge currents are modified by the presence of the fermions as follows
 \ref\KASU{Y. Kazama and H. Suzuki, Nucl. Phys. {\bf B234} (1989) 232 \semi
Phys. Lett. {\bf 216 B} (1989) 112.}

\eqn\one{J_3 \to J_3 '=J_3+\psi_+ \psi_- \ ,\qq
I_2 \to I_2 '=I_2 +\chi_+ \chi_-\ ,}
where $\psi_{\pm}=\psi_2 \pm \psi_1$ and $\chi_{\pm}=\chi_3 \pm i \chi_1$.
Next we construct the gauge currents for the heterotic string

\eqn\six{\eqalign{&Q_{het}=J_3 '-\sin \a\ \bar J_3+\sqrt{k'\over k}\cos\beta
 \bar I_2 \cr
&\tilde Q_{het}=-\sqrt{k'-2\over k+2} I_2 ' -\cos \a \bar J_3 -\sqrt{k'\over k}
\sin \a \bar I_2\ .\cr } }
and the gauge currents for the type-II string

\eqn\three{\eqalign{
&Q_{II}=J_3 '-\sin \a \bar J_3 '+\sqrt{k'-2\over k+2}\bar I_2 ' \cr
&\tilde Q_{II}=-\sqrt{k'-2\over k+2} I_2 ' -\cos \a \bar J_3 '
-\sqrt{k'-2\over k+2} \sin \a \bar I_2 '\ .} }
Note the shifts in the $k,k'$ that are quite different from each other and
from the bosonic case in \qqs . These shifts are required in the quantum
theory in order to maintain the gauge invariance of the stress tensor in each
case. This can be verified by demanding that the operator products of the
stress tensor and the gauge currents vanish. The stress tensor for the left
movers is

\eqn\four{T=T_G +T_{\psi\chi} -{(Q_L)^2 \over k'-2}
-{(\tilde Q_L)^2 \over k'-2}\ ,}
where $(Q_L,\tilde Q_L)$ are the left-moving pieces of the modified gauge
currents. Note that their central extensions have also shifted relative to the
bosonic case. Here $T_G$ is the standard Sugawara stress tensor for the group
$SL(2,\IR)_{-k'}\otimes SU(2)_k$ and $T_{\psi\chi}$ is the stress tensor for
the free fermions $(\psi_1,\psi_2,\chi_1,\chi_3)$. The stress tensor for the
right movers will be like the bosonic case if we consider the heterotic
string, and it will be like above if we consider the type-II string.

Now that we have the exact $k,k'$ dependence of the operators we can compute
the transformations they generate on the group parameters. The trasformation
law will now be sensitive to the shifts in $k,k'$ and therefore the
construction of the invariants will change accordingly. The previous $r,s$
invariants remain as in \inva\ but $\rho,\lambda$ get modified. In the type-II
case we have

\eqn\five{
\r=\b+\g-{1\over 2}\sqrt{k'-2\over k+2}{\cos\a \over 1+\sin\a}\ln{a\over b},
\quad
\l=\g-\b+{1\over 2}\sqrt{k'-2\over k+2}{\cos\a \over 1-\sin\a} \ln{u\over v}
\ .}
while for the heterotic case we obtain

\eqn\seven{
\r=\b+\sqrt{k'(k+2)\over k(k'-2)} \g -{1\over 2}\sqrt{k'\over k}
{\cos \a \over 1+\sin \a}\ln {a\over b},\ \
\l=\sqrt{k'(k+2)\over k(k'-2)} \g -\b +{1\over 2}\sqrt{k'\over k}
{\cos \a \over 1-\sin \a} \ln{u\over v}\ .}
Using these expressions we can now repeat the procedure of \lzero\ to compute
the metric and dilaton. For the type-II case the result is quite simple: the
exact expressions coincide with the semi-classical expressions in \classic\
except for shifting $k'$ to $(k'-2)$ and $k$ to $(k+2)$. However, for the
heterotic string the result is considerably complicated and will not be given
here.

In this paper we did not discuss the antisymmetric tensor, or torsion field,
that is present for this coset model. In the algebraic approach it can be
computed by considering states that satisfy the closed string conditions
$L_0=\bar L_0$ without demanding that $Q=\tilde Q=0$ on the state. In the
coset approach such states must be included in the spectrum. What will then
happen is that the Dalambertian will have to include a ``spin connection" in
addition to the metric. The torsion field can then be extracted by comparing
the derivative form of $L_0+\bar L_0$ and the Dalambertian form. We have not
applied this method in detail to the current coset model.

In summary, we have computed the conformally exact
metric and dilaton and gave the global space for the curved spacetime
string model based on the coset $SL(2,\IR)\times SU(2)/\IR^2$ . We have also
discussed the supersymmetric type-II and heterotic versions of the model.
As seen, the finite $k,k'$ corrections are substantial not only numerically
but also structurally.

\listrefs


\end